\documentclass{css2018}

\usepackage{epsfig}

\title{Constraints on the existence \\ of dark matter haloes \\
         by the M81 group \\
         and the Hickson compact groups of galaxies}

\author{W. Oehm$^1$, P. Kroupa$^{2,3}$ \\
\vskip 2mm {\small
$^1$
Bonn, Germany\\
physik@wolfgang-oehm.com \\
$^2$
Helmholtz Institut f\"ur Strahlen und Kernphysik \\
Universit\"at Bonn, Nussallee 14 - 16, 53115 Bonn, Germany \\
pavel@astro.uni-bonn.de \\
$^3$
Astronomical Institute, Faculty of Mathematics and Physics, Charles University \\
V  Hole\v{s}ovi\v{c}k\'ach 2, CZ-180 00 Prague 8, Czech Republic \\
}}

\abstract{
According to the standard model of cosmology the visible, baryonic matter of galaxies
is embedded in dark matter haloes, thus extending the mass and the size of galaxies by  
one to two orders of magnitude. Taking into account dynamical friction between the dark matter haloes,
the nearby located M81 group of galaxies as well as the Hickson compact groups of galaxies are here
investigated with regard to their dynamical behaviour. The results of the employment of the Markov Chain Monte Carlo method and the genetic algorithm show statistically substantial merger rates between galaxies,
and long living constellations without merging galaxies comprise - apart from very few instances -
initially unbound systems only.
This result is derived based on three- and four-body calculations for
a model of rigid Navarro-Frenk-White profiles for the dark matter haloes, but verified by the
comparison to randomly chosen  individual solutions for the M81 galaxy group
with high-resolution simulations 
of live self-consistent systems ($N$-body calculations).
In consequence, the observed compact configurations of major galaxies are a very unlikely occurence
if dark matter haloes exist.
}


\begin{document}

\maketitle

\section{Introduction}  %----------------------------------------------------------------
\label{sec:1}

Radioastronomical observations 
(\cite{Cottrell-1977}, \cite{Gottesman-1977}, \cite{Hulst-1979}, 
\cite{Appleton-1981}, \cite{Yun-1993}, \cite{Yun-1994})
established the fact that the M81 companions  M82 and NGC~3077 are
connected with the central galaxy M81 by intergalactic clouds of $H_I$ emitting gas,
namely the north and the south tidal bridge.
Attempting to reproduce those morphological structures by numerical simulations,
based on the dark matter hypothesis underlying the cosmological model,
Yun \cite{Yun-1999} couldn't find solutions 
for the dynamic development of the inner M81 group of galaxies without the 
occurence of merging galaxies due to dynamical dissipation. 
Employing full $N$-body calculations, Thomson et al.~(\cite{Thomson-1999}) didn't
find appropriate solutions without a merger either.  Although more recent observational work exists, 
the dynamical evolution of the inner M81 group has not been investigated theoretically 
since then, until Oehm et al.~(\cite{Oehm-2017}) investigated the inner M81 group 
regarding their dynamical behaviour including the effects of dynamical friction between 
the dark matter haloes from a statistical point of view. The results obtained there
disfavour the existence of dark matter haloes according to the model,
and are summarised in Chapter~\ref{sec:3}. \\

Currently we transfer the methodology applied to the M81 Group 
to the Hickson compact groups of galaxies
(\cite{Hickson-1982}, \cite{Hickson-1988}, \cite{Hickson-1992}, \cite{Hickson-1997}),
based on investigations recently published by Sohn et al.~(\cite{Sohn-2015}).
Comparably to the M81 group, the preliminary results obtained for a subset of 100 
compact groups also disfavour the existence of dark matter haloes because of 
significant merger probabilities. The approach and the preliminary results are 
presented in Chapter~\ref{sec:4}. \\

At first, the underlying physical model for the statistical evaluations is explained in 
Chapter~\ref{sec:2}:

\section{The Model}   %----------------------------------------------------------------
\label{sec:2}

The DM halo of either galaxy is treated as a rigid halo with a density
profile according to Navarro, Frenk and White \cite{Navarro-1995} (NFW-profile), truncated at
the radius $R_{200}$:
\begin{equation}
\label{eq:NFW}
\rho (r)=\frac{\rho_0}{r/R_s\left(1+r/R_s\right)^2}\ ,
\end{equation}
with $R_s=R_{200}/c$, $R_{200}$ denoting the radius yielding an
average density of the halo of 200~times the cosmological critical
density
\begin{equation}
\rho_{crit}=\frac{3 H^2}{8 \pi G}\  ,
\end{equation}
and the concentration parameter~$c$
\begin{equation}
\log_{10}c=1.02-0.109\left(\log_{10}\frac{M_\mathrm{vir}}{10^{12}M_{\odot}}\right)\ 
\end{equation}
(see \cite{Maccio-2007}). \\

The DM~halo masses are derived from the luminosities of the galaxies available at the 
NASA/IPAC Extragalactic Database for the M81 group (query submitted on 2014 February 8),
and in \cite{Sohn-2015} for the Hickson compact groups. In a first step
the stellar masses are determined by means of eq.~6 of \cite{Bernardi-2010}, 
and based on the stellar masses the DM~halo masses are extracted 
from fig.~7 of \cite{Behroozi-2013} thereafter.  \\

Exploring the dynamics of bodies travelling along paths in the
interior of DM haloes implies that the effects of dynamical friction
have to be taken into account in an appropriate manner \cite{Chandra-1942}.
For isotropic distribution functions the deceleration of an
intruding point mass due to dynamical friction is described by
Chandrasekhar's formula \cite{Chandra-1943}, which reads for a
Maxwellian velocity distribution with dispersion $\sigma$ (for
details see \cite{Binney-2008}, chap. 8.1):
\begin{equation}
\label{eq:chandra}
\frac{d\vec{\textbf{v}}_M}{dt} = -\frac{4{\pi}G^2M{\rho}}{{\textnormal{v}_M}^3}\ \textnormal{ln}{\Lambda} 
\left[\textnormal{erf}(X) - \frac{2X}{\sqrt{{\pi}}}\textnormal{e}^{-X^2} \right] \vec{\textbf{v}}_M\ ,
\end{equation}
with \emph{X} = v$_M$/($\sqrt{2}\sigma$). 
The intruder of mass \emph{M} and relative velocity $\vec{\textbf{v}}_M$ is decelerated by $d\vec{\textbf{v}}_M$/\emph{dt} in the background density $\rho$ of the DM~halo.

Simulating galaxy-galaxy encounters Petsch and Theis (\cite{Petsch-2008}) showed that a
modified model for the Coulomb logarithm ln$\Lambda$, originally
proposed by Jiang et al. (\cite{Jiang-2008}), describes the effects of dynamical
friction in a realistic manner. This mass- and distance-dependent
model reads:
\begin{equation}
\label{eq:cl}
\textnormal{ln}\Lambda = \textnormal{ln}\left[1 + \frac{M_{halo}(r)}{M}\right]\  ,
\end{equation} 
where $M_{halo}(r)$ is the mass of the host dark matter halo within the
radial distance $r$ of the intruding point mass.

So far Eq.~\ref{eq:chandra} describes the dynamical friction of a point mass in halo~$i$
with a Maxwellian velocity distribution. The approach how to calculate the dynamical 
friction between two overlapping haloes $i$ and $j$ using NFW-profiles is described in detail 
in appendix~C of \cite{Oehm-2017}.

Chandrasekhar's formula only gives an estimate at hand. However,
for the sake of establishing statistical statements about merger rates between galaxies,
high-resolution simulations of live self-consistent systems presented
in \cite{Oehm-2017} (especially refer to figures~13 and~14)
confirm our approach of employing this
semi-analytical formula in our three- and four-body calculations. \\

The equations of motion and the numerical approach of their integration are presented
in appendix~C of \cite{Oehm-2017}. 

\section{M81 review}   %---------------------------------------------------------------- 
\label{sec:3}

We briefly present the methodology and the results obtained for the galaxy group M81 (\cite{Oehm-2017}). \\

The fact that the three core members, M81, M82 and NGC~3077, are enshrouded by 
intergalactic clouds of $H_I$ emitting gas (north and south tidal bridge) implies that either companion M82 
and NGC~3077 must have encountered the central galaxy M81 closely within the recent cosmological
past (for a review see \cite{Yun-1999}). 

The plane-of-sky coordinates, the line-of-sight velocities, and the DM-halo masses are at our disposal. 
However, the plane-of-sky velocities are unknown, and the radial 
(line-of-sight) distances are only roughly 
established. Therefore, within the reference frame of the central galaxy M81, we are confronted with six open 
parameters: The radial distances and the plane-of-sky velocities of the companions M82 and NGC~3077. 

The possible values of those open parameters were investigated from a statistical
point of view:

At first, calculating three-body orbits backwards up to $-7$~Gyr,
statistical populations for the open parameters
were generated by means of the Markov Chain Monte Carlo method (MCMC) 
and the genetic algorithm (GA). Following the results of  \cite{Yun-1999} we added,
additionally to the known initial
conditions at present, the rather general condition that
\begin{center}
\emph{both companions M82 and NGC~3077 encountered M81}  \\
\emph{within the recent 500~Myr at a pericentre distance below 30~kpc.}
\end{center}
Each three-body orbit of those statistical populations is fully determined by all
the known and the open parameters provided by either MCMC or GA.
Starting at time $-7$~Gyr and
calculating the corresponding three-body orbits forward in time up to
$+7$~Gyr, the behaviour of the inner group has been investigated
with respect to the question of possibly occurring mergers in the
future. \\

The details of applying the Metropolis-Hastings algorithm based on a
methodology proposed by Goodman and Weare (\cite{Goodman-2010}) 
for MCMC are presented in section~4 and appendix~D of \cite{Oehm-2017}, 
and, based on the proposal by Charbonneau (\cite{Charbonneau-1995}) for GA 
in section~5 of \cite{Oehm-2017}. Basically both methods 
deliver comparable results. However, as discussed in section~6 of \cite{Oehm-2017},
due to the structure of the likelihood function applied for MCMC in our case,
the genetic algorithm has been proved to deliver more stable results. \\

%%%%%%%%%%%%%%%%%%%%%%%%%%%%%%%%%%%
\begin{table*}           
\centering                        
\begin{tabular}{c c c}         
\hline\hline                  
                                                                                           & MCMC           & GA                   \\
\hline 
solutions not merging within next 7 Gyr                            & 118               &  278                 \\ 
\hline               
solutions not merging  within next 7 Gyr and:                   &                      &                         \\                     
neither M82 nor N3077 bound to M81 7 Gyr ago              & 117               &  276                 \\ 
\hline               
solutions not merging  within next 7 Gyr and:                   &                      &                         \\                     
one companion bound to M81 7 Gyr ago                           & 1                   &  2                    \\          
\hline           
solutions for:                                                                      &                       &                          \\                          
M82 and N3077 bound to M81 7 Gyr ago                          &  66                 &   70                  \\ 
\hline                                   
longest lifetime from today for:                                         &                       &                          \\ 
M82 and N3077 bound to M81  7 Gyr ago                         &  $2.7$ Gyr     &   $2.8$ Gyr      \\ 
\hline                                   
average lifetime from today for:                                        &                       &                          \\ 
M82 and N3077 bound to M81  7 Gyr ago                         &  $1.7$ Gyr      &   $1.3$ Gyr       \\ 
\hline
\end{tabular}
\caption{\emph{Galaxy group M81:}
  Key numbers for both statistical methods MCMC and GA, based
  on populations of 1000 solutions in either case. Actually, the three
  solutions not merging within the next 7~Gyr where one companion is
  bound to M81 (third position)
  merge after 7.3~Gyr (MCMC), and 7.8 and 8.2~Gyr (GA).}
\label{tab:M81}      
\end{table*}                
%%%%%%%%%%%%%%%%%%%%%%%%%%%%%%%%%%%

In Table~\ref{tab:M81} (which is table~4 in \cite{Oehm-2017}) we present the basic 
results of our statistical evaluations which can be summarised as follows:

\begin{itemize}

\item{Long living solutions without mergers comprise constellations only where the three 
  galaxies are unbound and - arriving from a far distance - happen to simultaneously encounter
  each other within the previous 500~Myr.
}

\item{Cases where all three galaxies are bound at $-7$~Gyr represent only $7\%$
  of either statistical population of the 
  MCMC and GA solutions, respectively. And 
  those originally bound systems would be merging within the near cosmological future.
}

\end{itemize}

\section{The Hickson Compact Groups}   %----------------------------------------------------------------
\label{sec:4}

Upon having established our methods for the galaxy group M81 we transferred this methodology 
to the Hickson compact groups 
with three and four members based on the observational data summarised by Sohn et al. (\cite{Sohn-2015}).
The aim is to achieve statistical results for the merger rates for a subset of 188 compact groups extracted from  
the list of originally 332 compact groups presented in \cite{Sohn-2015}. 
The reasons for the non-consideration of 144 groups are:

\begin{itemize}

\item{There are groups where the true membership of at least one galaxy is not clarified (52 cases).}

\item{We don't consider groups consisting of more than four true members (28 cases).}

\item{Some groups are omitted due to only inaccurately known line-of-sight velocities 
  (spread of redshifts, $\Delta z = 0.001$ being too large) (9 cases).}

\item{Groups consisting of galaxies with a DM-halo mass exceeding $10^{15}M_\odot$ 
  are not taken into account because the determination of the DM-halo masses based 
  on the stellar masses according to Behroozi et al. (\cite{Behroozi-2013}) is confined to 
  the interval $\left[10^{10}M_\odot , 10^{15}M_\odot \right]$ for the DM-halo masses (94 cases).}

\end{itemize}

Of course two ore more criteria can simultaneously 
apply to one compact group, therefore ending up with 188 groups 
to be considered, which is $57\%$ of the original set.
However, as the investigations were still in process when creating this article, the
preliminary results presented here are based on a subset of the nearest 100 compact groups 
(see Appendix~A) from this set of 188 objects. 
The range of distances for our preliminary set of 100 groups is 
$\left[65~\rm{Mpc} , 308~\rm{Mpc} \right]$. \\

The average plane-of-sky distance between two galaxies for our final set of 188 compact groups 
is $92.7~\rm{kpc}$. The assumption of  isotropy yields an average spatial distance between two galaxies 
of $113~\rm{kpc}$. Physical intuition already implies that dynamical friction between DM-haloes
with radii of hundreds of $\rm{kpc}$ plays an important role regarding the dynamical
behaviour of the groups.

Following our methodology established for the M81 group, we specify the general condition for 
the compact groups that
\begin{center}
\emph{the minimal value for the hyper radius does not exceed a certain ceiling value} \\ 
\emph{within the recent Gyr, i.e. $\left[-1~Gyr, today \right]$.}
\end{center}
To be precise, the hyper radius,~$\rho$, defined by
\begin{equation}
\label{eq:hyper-1}
\rho^2 = \   \sum_{i < j}^{n} \  r_{ij}^2 \ \ ,  
\end{equation} 
for the $n$~members of each group shall fullfil the following condition for the minimal value~$\rho_{min}$
\begin{equation}
\label{eq:hyper-2}
\rho_{min}^2 < \ \rho_0^2 \  = \ \frac{n (n - 1)}{2} \ r_0^2 \ \ \ \
\textnormal{within} \ \left[ -1 \ \textnormal{Gyr, today} \right] \  
\end{equation} 
with ceiling value~$\rho_0$. Concerning~$r_0$ we consider three different values:
\begin{equation}
\label{eq:hyper-3}              
r_0 = \left\{
\begin{array}{ll}
\ 75 \ kpc &  \textnormal{(model~$A$)},  \\
100 \ kpc &  \textnormal{(model~$B$)},  \\
113 \ kpc &  \textnormal{(model~$C$)},
\end{array} \right. 
\end{equation}
the models~$A$ and~$B$ being motivated by the statement that the compact groups 
have recently gone through a configuration where the average distance between the 
individual members doesn't exceed two to three times the value of the visible, 
baryonic diameter of the galaxies. Model~$C$ refers to the above mentioned observed average
spatial distance between two galaxies within a group. \\

Employing the genetic algorithm we confine, within the reference frame of the most massive
galaxy of each group, 
for the remaining members the hardly known line-of-sight distances to 
$\left[-1~\rm{Mpc}, +1~\rm{Mpc} \right]$, 
and the unknown plane-of-sky Cartesian velocity components to realistic values of 
$\left[-500~\rm{pc/Myr}, +500~\rm{pc/Myr} \right]$. 
The range for the line-of-sight distances is obviously justified by the average spatial distance
of~$113~\rm{kpc}$ between galaxies, and
the constraints for the velocity components are actually in agreement with the results obtained
by Sohn et al. (\cite{Sohn-2015}, figures~5 and~6). 

As a matter of fact, according to Hickson~(\cite{Hickson-1997}) and Sohn et al.
this choice of ranges for the unknown entities could be confined even more drastically. 
However, in order not to influence the statistics by "wishful" a priori constraints, 
we take our decision for this choice of ranges.

The fitness function - which corresponds to the likelihood function of MCMC -
is defined by referring to the ceiling values~$\rho_0$ from Eq.~\ref{eq:hyper-2}
for the various models of Eq.~\ref{eq:hyper-3}:
\begin{equation}
\label{eq:fit-E}              
f(\rho_{min}) = \left\{
\begin{array}{ll}
1, 
&  \rho_{min} \le \rho_0   \ , \\
\exp\left({-{\displaystyle\frac{\left( \rho_{min}-\rho_0 \right)^2}{2\cdot {(25~kpc)}^2}}}\right), 
&  \rho_{min} > \rho_0 \ .
\end{array} \right. 
\end{equation}            \\

\begin{figure*}
\centering
\begin{tabular}{cc}
\includegraphics[width=8cm]{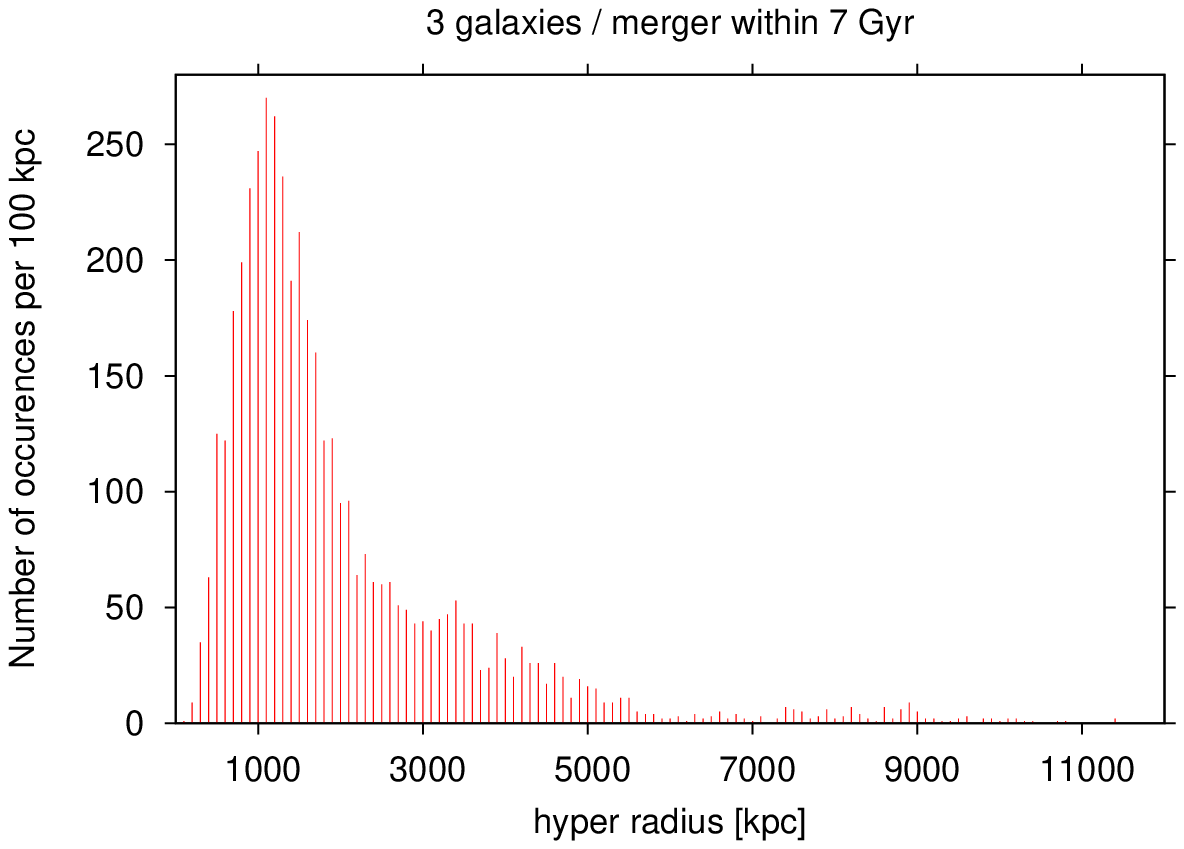} & \includegraphics[width=8cm]{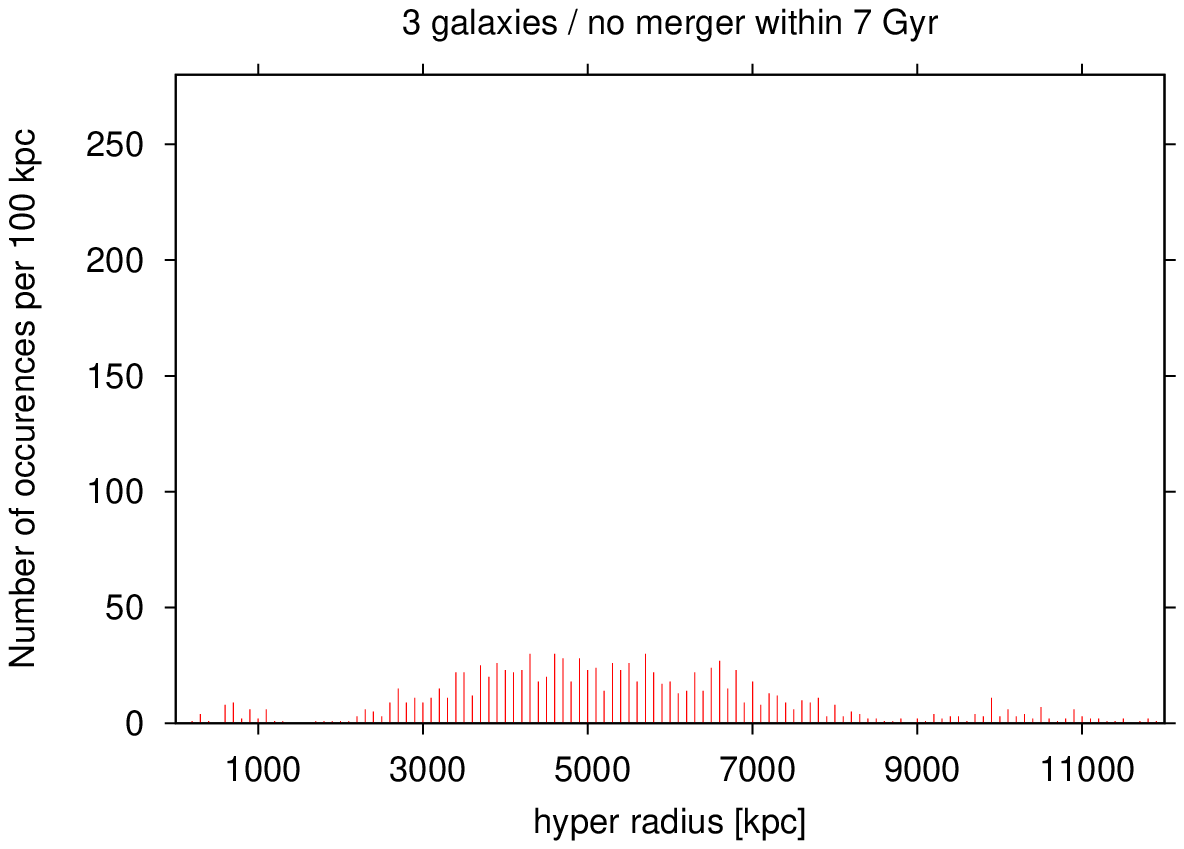} \\    % <=====
\includegraphics[width=8cm]{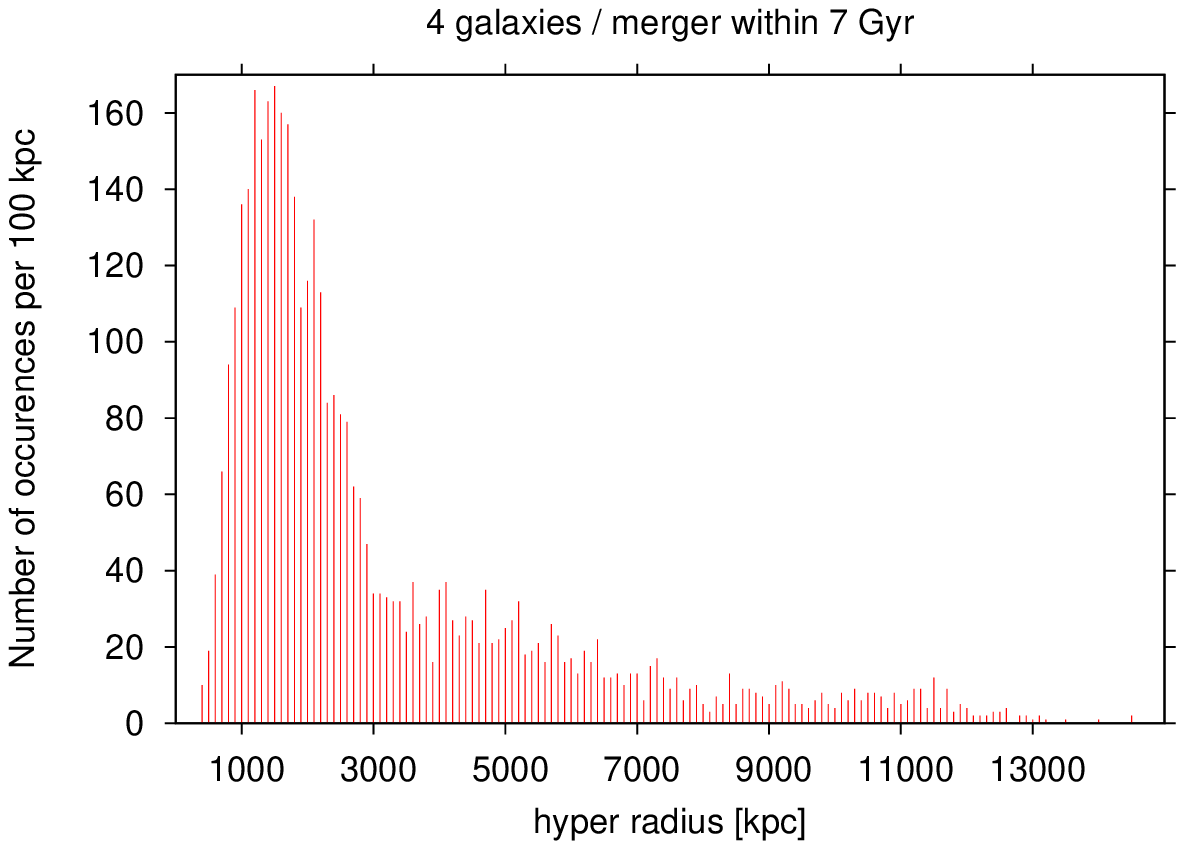} & \includegraphics[width=8cm]{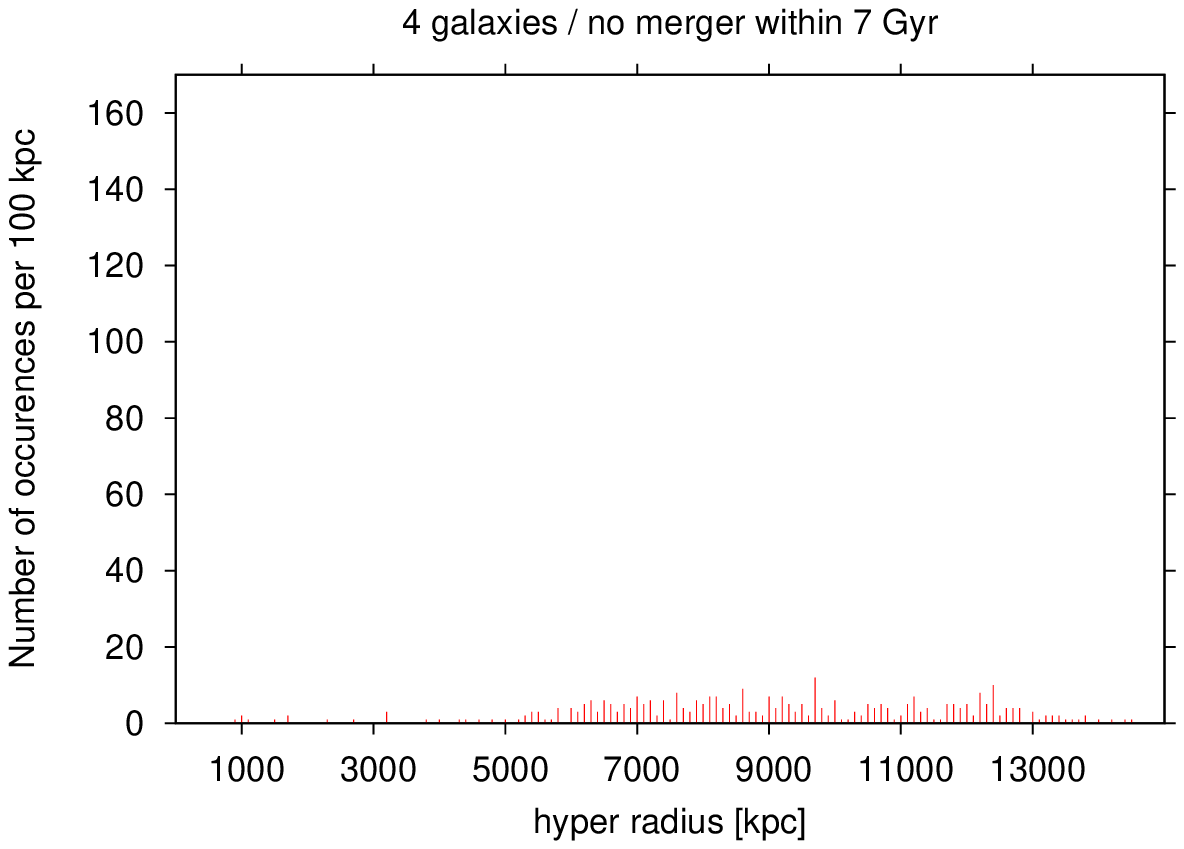} \\    % <=====
\end{tabular}
\caption{\emph{Compact groups:} Hyper radii at -7~Gyr for the three-galaxy groups
  (top panel) and the four-galaxy groups (bottom panel). 
  The left panel shows the number of occurrences for intervals of 10~kpc for solutions
  where at least one pair of galaxies will be merging within the next 7~Gyr, while the right
  panel refers to solutions without a merger within the next 7~Gyr.
  The data are extracted from the results for Model~$C$.}
\label{fig:hr}
\end{figure*}

\begin{figure*}
\centering
\begin{tabular}{cc}
\includegraphics[width=8cm]{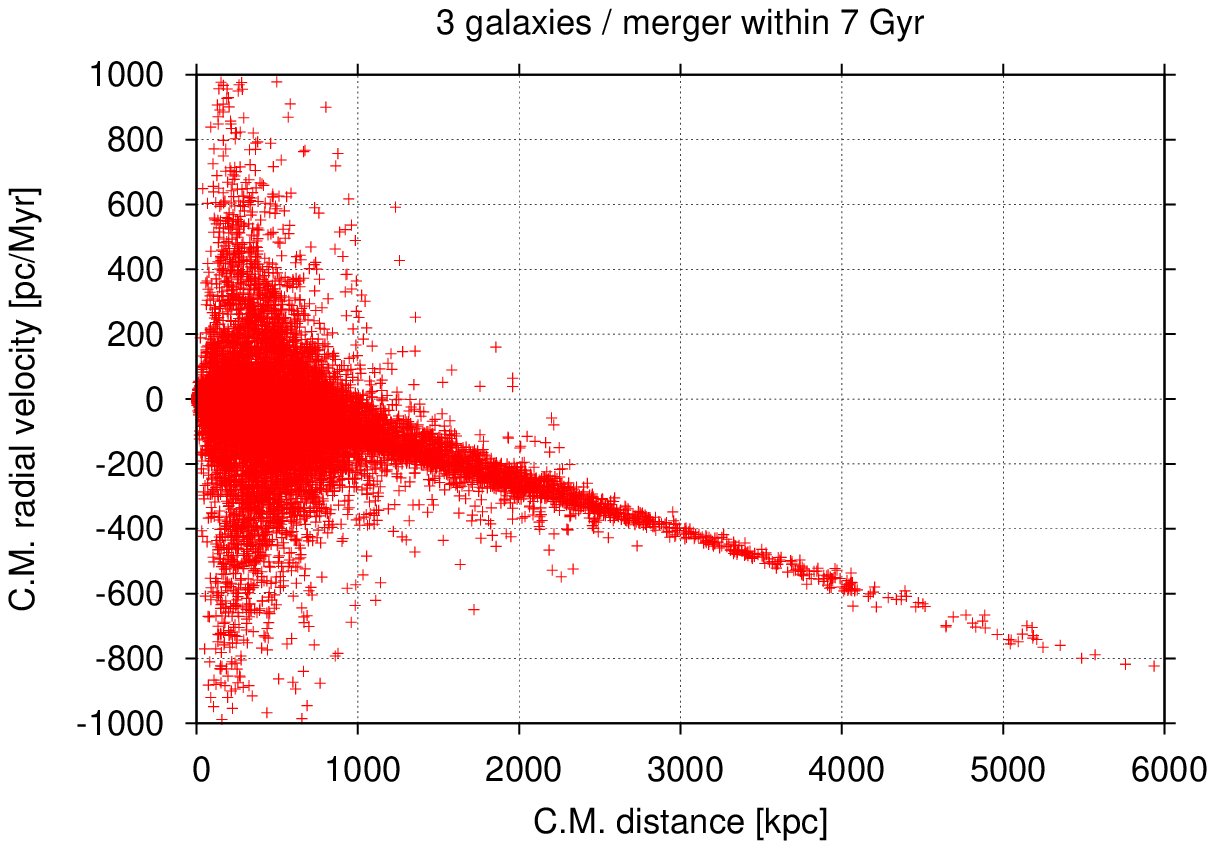} & \includegraphics[width=8cm]{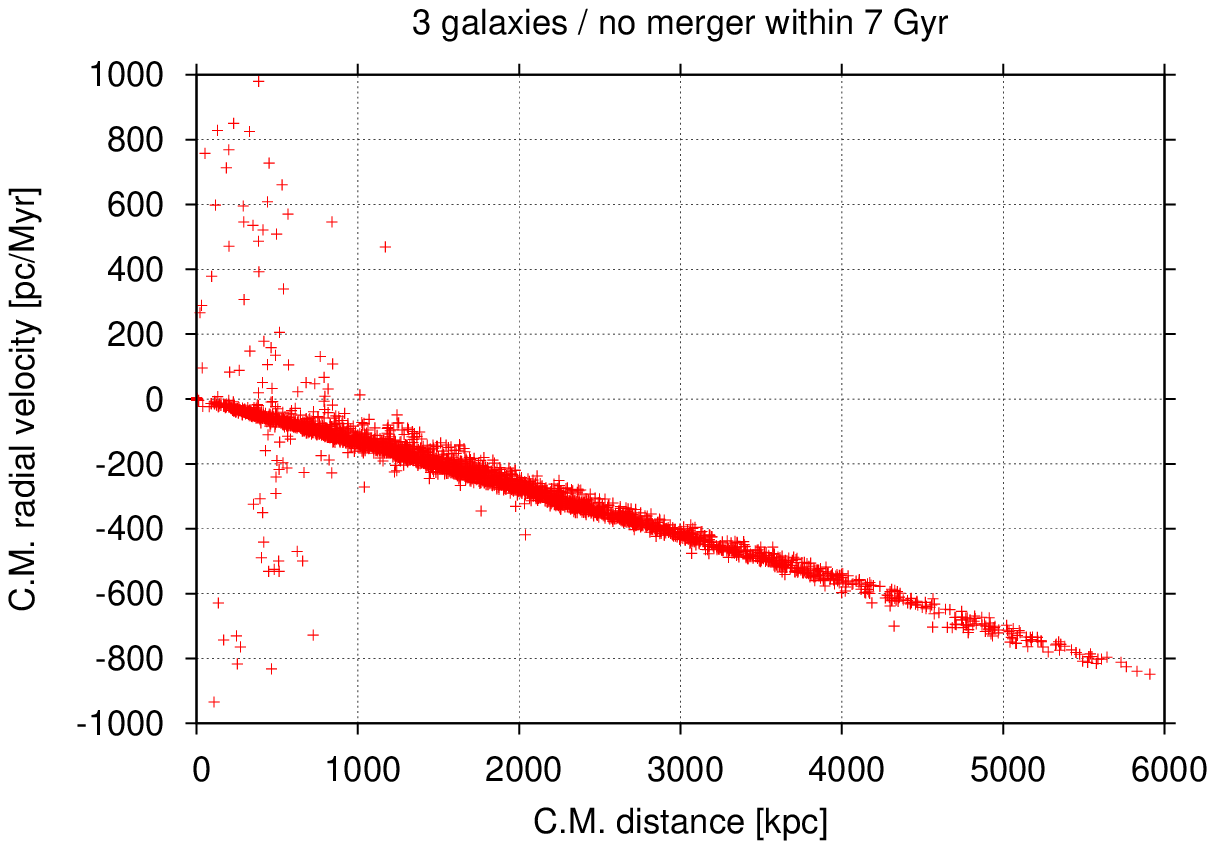} \\    % <=====
\includegraphics[width=8cm]{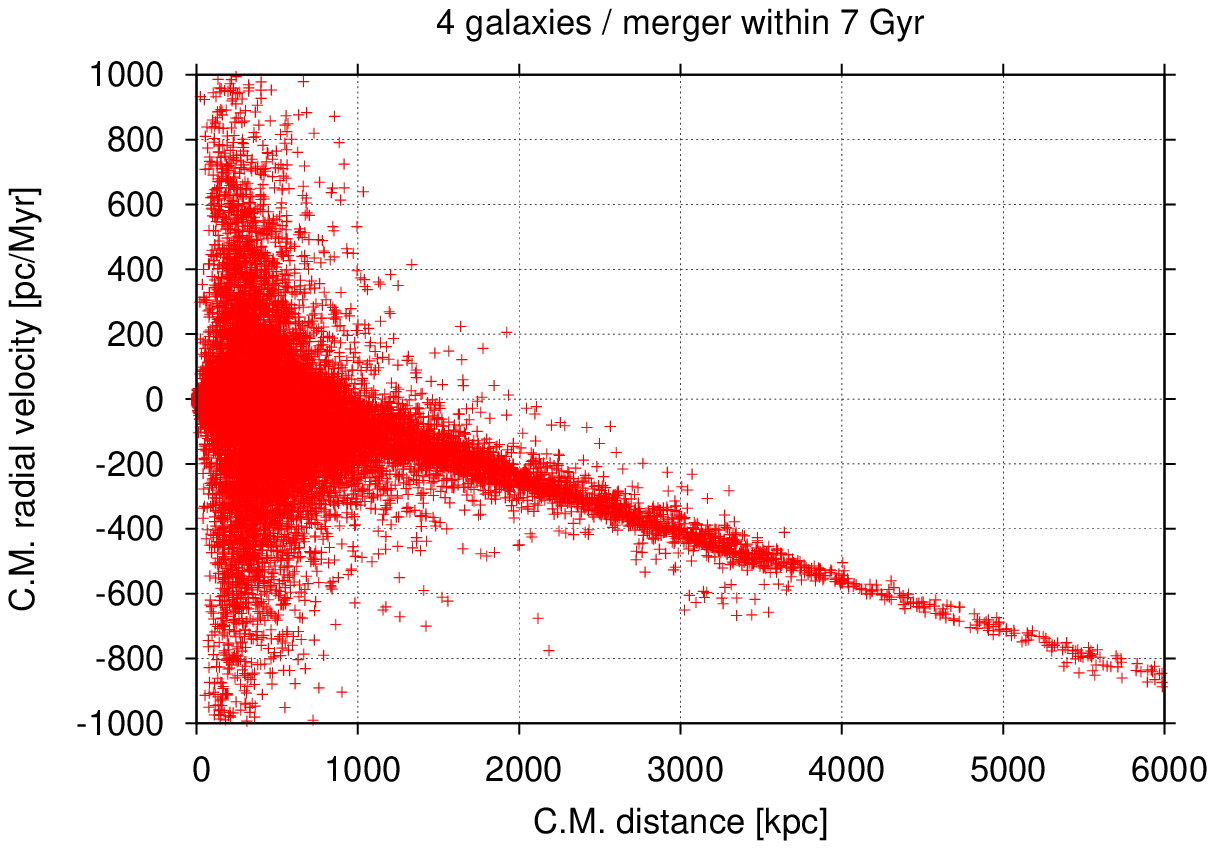} & \includegraphics[width=8cm]{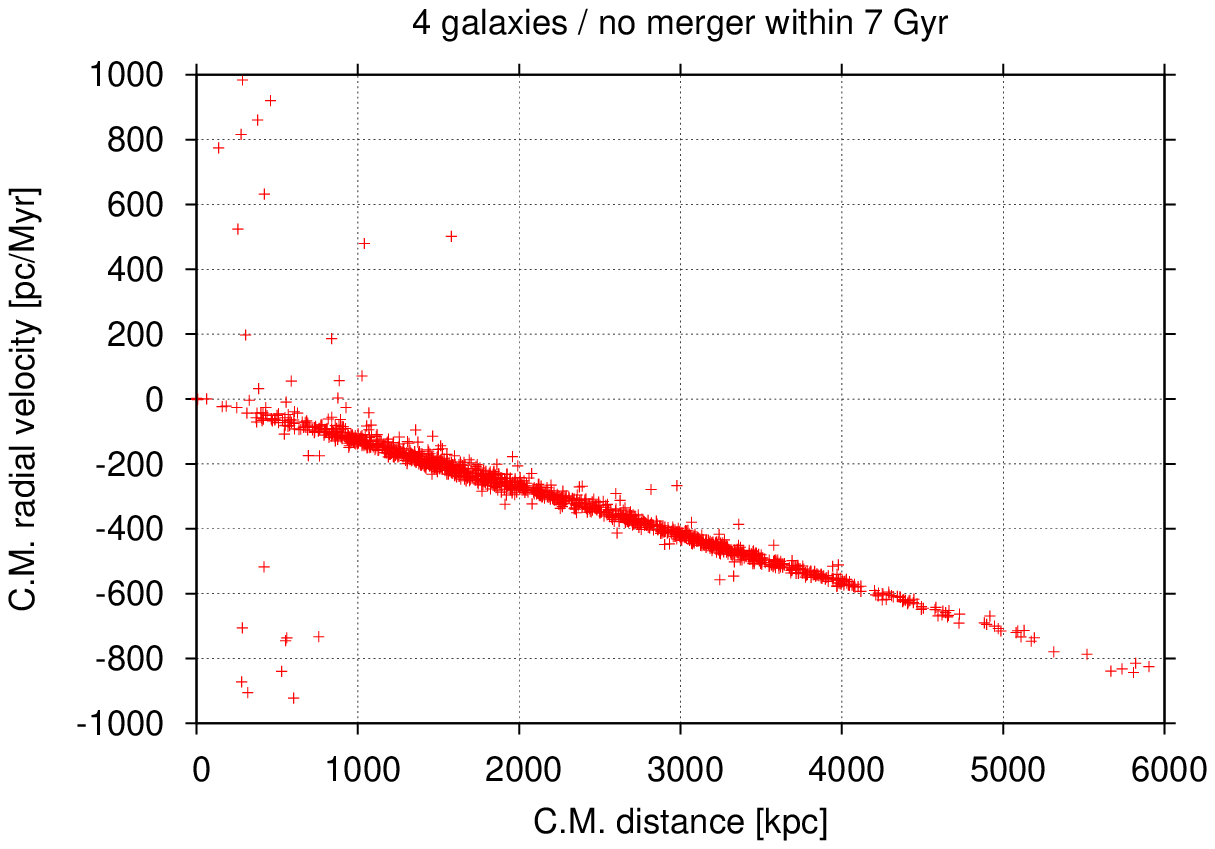} \\    % <=====
\end{tabular}
\caption{\emph{Compact groups:} The centre of mass radial velocities of the galaxies 
  in dependence of the centre of mass distances
  at -7~Gyr for the three-galaxy groups (top panel) and the four-galaxy groups (bottom panel). 
  Similar to Fig.~\ref{fig:hr}, the left panel refers to solutions
  where at least one pair of galaxies will be merging within the next 7~Gyr, while the right
  panel refers to solutions without a merger within the next 7~Gyr.
  The data are extracted from the results for Model~$C$.}
\label{fig:rv}
\end{figure*}

The results obtained are presented in Tables~\ref{tab:merger-2} and~\ref{tab:merger-tot} as well 
as in Figures~\ref{fig:hr} and~\ref{fig:rv}.

\begin{itemize}

\item{Although the models $A$, $B$, and $C$ cover a substantial variation of~$r_0$, the long
  term merger percentages show absolutely comparable numbers 
  (see Tables~\ref{tab:merger-2} and~\ref{tab:merger-tot}).
  Each model delivers the result that more than half of the considered preliminary set of 
  compact groups will totally be merged to one galaxy within the next 7~Gyr (Table~\ref{tab:merger-tot}),
  as well as that for about $2/3$ of the groups at least one pair of galaxies will be merging within 
  the next 2~Gyr (Table~\ref{tab:merger-2}). 
  The major difference between the models is just a slight delay for merging galaxies
  caused by a higher value of~$r_0$, thus above all affecting the percentages for 0-1~Gyr.
}

\item{Apart from very few exemptions, the hyper radii at -7~Gyr 
  shown in Fig.~\ref{fig:hr} clearly indicate
  that non-merging long living solutions comprise unbound systems only. Actually, instances with 
  hyper radii less than 1~Mpc for the three-galaxy groups concern only three out of 56 compact 
  groups, namely SDSSCGA00488, 01220, and 01446. For the four-galaxy groups only two 
  out of 44 compact groups comprise solutions with hyper radii less than 2~Mpc, namely 
  SDSSCGA00425 and 00800. Emphasising our statement, Fig.~\ref{fig:rv} shows that 
  the non-merging solutions almost completely comprise instances where the individual galaxies are, 
  at -7~Gyr, moving towards the centre of mass of their group with linearly increasing radial
  velocities in dependence of the centre of mass distance. 
}

\end{itemize}

%%%%%%%%%%%%%%%%%%%%%%%%%%%%%%%%%%%
\begin{table*}           
\centering                        
\begin{tabular}{c c c c c c c c}  
& & & & & & & \\   
\hline\hline                  
Model                     & 0-1 Gyr   & 0-2 Gyr    & 0-3 Gyr    & 0-4 Gyr   & 0-5 Gyr   & 0-6 Gyr    & 0-7 Gyr       \\ 
\hline            
$A$                        & 38\%      & 68\%       & 79\%        & 83\%     & 86\%      & 87\%       & 88\%      \\
$B$                         & 31\%     & 63\%       & 76\%        & 81\%     & 84\%      & 85\%       & 86\%      \\
$C$                        & 28\%      & 61\%       & 74\%       & 79\%      & 82\%      & 84\%       & 85\%       \\
\hline
$C-3$                     & 23\%     & 53\%       & 66\%        & 73\%     & 76\%       & 78\%      & 80\%      \\
$C-4$                     & 36\%     & 72\%       & 84\%        & 88\%     & 90\%       & 91\%      & 92\%      \\   
\hline
\end{tabular}
\caption{\emph{Compact groups:} Percentages of mergers for selected periods of time
  from the present until maximally 7~Gyr, cumulated over our 
  preliminary set of 100 selected groups. The numbers refer to the first occurence
  of a merging pair of galaxies for each group and are based on populations of 100 solutions
  per group, yielding in total a set of 10.000 solutions.
  The first three rows show the total percentages for the three models of the fitness function, 
  while the fourth and the fifth row refer to 56 three-galaxy groups and 44 four-galaxy groups,
  respectively, for model~$C$.}
\label{tab:merger-2}      
\end{table*}                
%%%%%%%%%%%%%%%%%%%%%%%%%%%%%%%%%%%

%%%%%%%%%%%%%%%%%%%%%%%%%%%%%%%%%%%
\begin{table*}           
\centering                        
\begin{tabular}{c c c c c c c c}  
& & & & & & & \\   
\hline\hline                  
Model                     & 0-1 Gyr   & 0-2 Gyr    & 0-3 Gyr    & 0-4 Gyr   & 0-5 Gyr   & 0-6 Gyr    & 0-7 Gyr       \\ 
\hline            
$A$                        & 4\%      & 24\%       & 37\%        & 45\%     & 51\%      & 55\%       & 58\%      \\
$B$                         & 3\%     & 21\%       & 33\%        & 41\%     & 46\%      & 50\%       & 54\%      \\
$C$                        & 2\%      & 19\%       & 31\%        & 39\%     & 44\%      & 49\%       & 53\%       \\
\hline
$C-3$                     & 3\%     & 23\%       & 34\%        & 41\%     & 46\%       & 50\%      & 53\%      \\
$C-4$                     & 1\%     & 14\%       & 26\%        & 36\%     & 42\%       & 48\%      & 52\%      \\   
\hline
\end{tabular}
\caption{\emph{Compact groups:} Same as for Table~\ref{tab:merger-2},
  but the numbers refer to complete mergers
  (all individual galaxies of a group will have merged to one object).}
\label{tab:merger-tot}      
\end{table*}                
%%%%%%%%%%%%%%%%%%%%%%%%%%%%%%%%%%%

%%%%%%%%%%%%%%%%%%%%%%%%%%%%%%%%%%%
%%%%%%%%%%%%%%%%%%%%%%%%%%%%%%%%%%%

%%%%%%%%%%%%%%%%%%%%%%%%%%%%%%%%%%%
%%%%%%%%%%%%%%%%%%%%%%%%%%%%%%%%%%%

\section{Conclusions}   %---------------------------------------------------------------- 
\label{sec:5}

We apprehend the statistically elaborated merger percentages for the M81 group and the 
Hickson compact groups of galaxies as a \emph{merger probability per time unit} 
for those systems. 
The solutions of the configuration of these groups 7~Gyr ago obtained under the condition
that the groups have not merged by the present time comprise virtually only cases where 
the galaxies making-up the present-day groups are moving towards each other from 
large distances ($>~1$~Mpc). It appears unlikely for this correlated motion to be realistic.

\section*{Appendix}  %----------------------------------------------------------------  

The 100 compact groups from the \emph{SDSS}~catalogue considered 
in this publication are (56 groups with three and 44 with four members): \\

SDSSCGA00027,
00029	,
00037	,
00110	,
00113	,
00131	,
00132	,
00177	,
00240	,
00275	,
00309	,
00345	,
00355	,
00375	,
00397	,
00407	,
00418	,
00425	,
00435	,
00483	,
00488	,
00510	,
00539	,
00621	,
00630	,
00673	,
00676	,
00711	,
00728	,
00735	,
00736	,
00752	,
00755	,
00798	,
00800	,
00811	,
00820	,
00895	,
00902	,
00912	,
00916	,
00933	,
00934	,
00954	,
00960	,
01012	,
01020	,
01036	,
01056	,
01059	,
01065	,
01076	,
01136	,
01139	,
01184	,
01220	,
01244	,
01251	,
01264	,
01265	,
01277	,
01300	,
01303	,
01327	,
01344	,
01372	,
01383	,
01391	,
01428	,
01434	,
01446	,
01458	,
01485	,
01503	,
01512	,
01528	,
01557	,
01563	,
01568	,
01605	,
01616	,
01667	,
01713	,
01724	,
01784	,
01841	,
01874	,
01932	,
02022	,
02037	,
02056	,
02188	,
02191	,
02192	,
02206	,
02237	,
02257	,
02270	,
02273	,
02277	.

\section*{Acknowledgements}  %----------------------------------------------------------------  

W. Oehm would like to express his gratitude for the support of
{scdsoft~AG} in providing a SAP system environment for the numerical
calculations. Without the support of {scdsoft's} executives
\emph{P. Pfeifer} and \emph{U. Temmer} the innovative approach of
programming the numerical tasks in SAP's language ABAP wold not have
been possible.

\end{document}